\documentclass[aps,twocolumn,letterpaper,prl,showpacs]{revtex4-1}
\usepackage{eurosym}
\usepackage{graphicx}
\usepackage{dcolumn}
\usepackage{bm}
\usepackage{color}
\usepackage{latexsym}
\usepackage{amssymb}
\usepackage{amsmath}

\begin{document}

\title{Slowing-down of non-equilibrium concentration fluctuations in confinement}
\author{C\'edric Giraudet$^{1}$, Henri Bataller$^{1}$, Yifei Sun$^{2}$, Aleksandar Donev$^{2}$, Jos\'e Maria Ortiz de Z\'arate$^{3}$ and Fabrizio Croccolo$^{1}$}

\address{$^{1}$Laboratoire des Fluides Complexes et leurs R\'eservoirs, Universit\'e de Pau et des Pays de l'Adour, 64600 Anglet, France}
\address{$^{2}$Courant Institute of Mathematical Sciences, New York University, New York, NY 10012, USA}
\address{$^{3}$Departamento de F\'{\i}sica Aplicada I, Universidad Complutense, 28040 Madrid, Spain}

\date{\today}

\begin{abstract}
Fluctuations in a fluid are strongly affected by the presence of a macroscopic gradient making them long-ranged and enhancing their amplitude. While small-scale fluctuations exhibit diffusive lifetimes, larger-scale fluctuations live shorter because of gravity, as theoretically and experimentally well-known. We explore here fluctuations of even larger size, comparable to the extent of the system in the direction of the gradient, and find experimental evidence of a dramatic slowing-down in their dynamics.
We recover diffusive behaviour for these strongly-confined fluctuations, but with a diffusion coefficient that depends on the solutal Rayleigh number. Results from dynamic shadowgraph experiments are complemented by theoretical calculations and numerical simulations based on fluctuating hydrodynamics, and excellent agreement is found. The study of the dynamics of non-equilibrium fluctuations allows to probe and measure the competition of physical processes such as diffusion, buoyancy and confinement.

\end{abstract}

\maketitle


It is well established that fluctuations are long-ranged in systems out-of-equilibrium~\cite{K01,K02,book_dezarate_2006}, even far from critical points where the long-range behaviour is observed also in equilibrium conditions~\cite{sengers_1986}. In a binary fluid mixture subject to a stabilizing (vertical) temperature or concentration gradient, the coupling between the spontaneous velocity fluctuations and the macroscopic gradient results in giant concentration fluctuations in the quiescent state~\cite{book_dezarate_2006,vailati_1997}. Gravity quenches the intensity of fluctuations with  length scales larger than a characteristic (horizontal) size $2\pi/q_s^\star $ related to the dimensionless solutal Rayleigh number $Ra_s$ of the system~\cite{vailati_1997,vailati_1998}:
\begin{align}
Ra_s&=\frac{\beta_s g \nabla c L^4}{\nu D};& -Ra_s=&(q_s^\star L)^4,
\label{eq_rayleigh}
\end{align}
where $\beta_s=\rho^{-1}(\partial \rho/\partial c)$ is the solutal expansion coefficient, $\rho$ the fluid density, $g$ the gravity acceleration, $c$ the concentration (mass fraction) of the denser component of the fluid, $\nabla c$ the modulus of the concentration gradient, $D$ the mass diffusion coefficient, $\nu$ the kinematic viscosity, and $q_s^\star $ a characteristic solutal wave vector.
Vertical boundaries suppress fluctuations larger than the confinement length $L$ in the direction of the gradient \cite{book_dezarate_2006,dezarate_2006}.
Gravity also accelerates the dynamics of the fluctuations for wavenumbers smaller than $q_s^\star $ via buoyancy effects, leading to non-diffusive decay of large-scale fluctuations~\cite{croccolo_2007}.

\begin{figure}[thb]
\center\includegraphics[width=8cm]{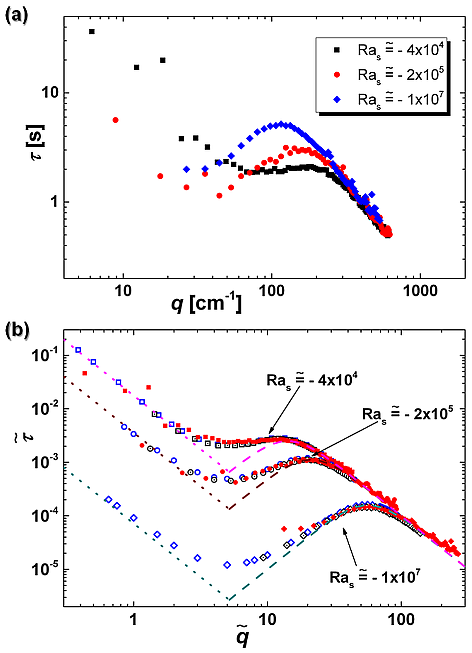}
\caption{Effective decay times: (a) Log-log plot of the experimental decay times $\tau$ as a function of wave vector $q$ for different Rayleigh numbers.
(b) Log-log plot of the dimensionless decay times $\tilde{\tau}$ as a function of dimensionless wave number $\tilde{q}$. Filled red markers are experimental data, open blue are for calculations based on the FHD model, and open-dotted black are from numerical simulations.
Dashed lines depict the analytical solution provided by Eq.~\eqref{eq_tau_bell} for $\tilde{q}>\tilde{q_b}$, taking into account gravity and diffusion only; dotted lines are for the confinement limit of Eq.~\eqref{eq_tau_conf} for $\tilde{q}<\tilde{q_b}$.}
\label{fig times}
\end{figure}

The dynamics of concentration non-equilibrium fluctuations (c-NEFs) in the presence of a vertical concentration gradient in a binary liquid mixture can be characterized in terms of the Intermediate Scattering Function (ISF or, equivalently, normalized time correlation function) $f(q,t)$, with $f(q,0)=1$.  At first approximation the ISF can be modeled by a single exponential with decay time $\tau(q)$ depending on the analysed wave vector $q$. Available theories accounting for the simultaneous presence of diffusion (d) and gravity (g)~\cite{segre_1993a,segre_1993b}, but not for confinement, predict for a stable configuration $(Ra_s<0)$:
\begin{equation}
\frac{\tau(\tilde{q})}{\tau_s} \bigg|_{d+g} =\tilde{\tau}(\tilde{q})|_{d+g}=\frac{1}{\tilde{q}^2\Bigg(1-\dfrac{Ra_s}{\tilde{q}^4}\Bigg)},
\label{eq_tau_bell}
\end{equation}
where the wave vector is expressed in its dimensionless form $\tilde{q}=qL$ and $\tau_s=L^2/D$ is the typical solutal time it takes diffusion to traverse the thickness of the sample. Equation~\eqref{eq_tau_bell} implies different behaviours for the decay times of small-scale and large-scale fluctuations,
$\tilde{\tau}(\tilde{q})|_{d} =1/\tilde{q}^2\quad \text{for}\quad  \tilde{q}\gg \tilde{q}_s^\star$,
and
$\tilde{\tau}(\tilde{q})|_{g} =-\tilde{q}^2/Ra_s\quad \text{for}\quad  \tilde{q}\ll \tilde{q}_s^\star$.
Actually, small fluctuations are dominated by diffusion, the latter being faster at small scales, while for large fluctuations buoyancy becomes more efficient and dominates the temporal evolution of c-NEFs. As a consequence, the fluctuation decay time has a maximum (clearly visible in the dashed lines of Fig.~\ref{fig times}) at $\tilde{q}_s^\star $, which identifies the most persistent fluctuation in the system if confinement is neglected.

The behaviour predicted by Eq.~\eqref{eq_tau_bell} has been experimentally verified in a number of experiments on c-NEFs related to a pure concentration gradient (isothermal mass diffusion)~\cite{croccolo_2007,croccolo_2006} or to a concentration gradient induced by the Soret effect~\cite{croccolo_2012,giraudet_2014,croccolo_2014}.

\begin{figure}[t]
\centering\includegraphics[width=8cm]{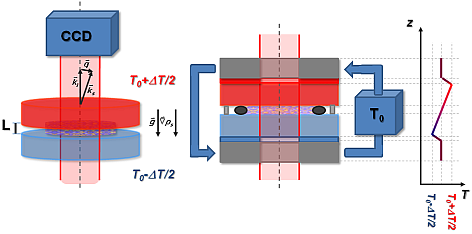}
\caption{Scheme of the experimental cell: two sapphire windows are kept at different temperatures $T_0+\Delta T/2$ (the top, red one) and $T_0-\Delta T/2$ (the bottom, blue one) while the sample fluid (colored pattern) is contained by an O-ring (black circles) at a thickness $L$ precisely defined by three plastic spacers (gray rectangles).}
\label{fig exp setup}
\end{figure}

Confinement is expected to cause deviations from Eq.~\eqref{eq_tau_bell} at very small wave numbers; to investigate this issue we perform experiments at wave vectors down to $q_{\text{min}}=8.9$~cm$^{-1}$. We apply a stabilizing temperature difference $\Delta T=20$ K (with an average temperature of $T_0=$298 K) to a horizontal layer of tetralin and n-dodecane at 50\% weight fraction of different vertical thicknesses $L=$ 0.7, 1.3 and 5.0 mm and constant lateral extent $R=$ 13.0 mm. The sample thickness is varied  by using different plastic spacers and sealing O-rings. Given the sample thermophysical properties~\cite{properties} the solutal Rayleigh numbers are: $Ra_s=-4\cdot10^4$, $-2\cdot10^5$ and $-1\cdot10^7$, respectively.  The thermal gradient cell is sketched in Fig.~\ref{fig exp setup}: two sapphire windows kept at fixed distance vertically contain the sample fluid and are thermally controlled by two Peltier elements with a central hole. The entire system allows a quasi-mono-chromatic parallel light beam pass through in the direction of the temperature gradient. More details of the thermal gradient cell can be found in previous literature~\cite{croccolo_2012,croccolo_2013}.

The rapid imposition of a temperature difference by heating the fluid mixture from above results in a linear temperature profile across the sample in a thermal time $\tau_T=L^2/\kappa$, where $\kappa$ is the fluid thermal diffusivity. Due to the much smaller value of the mass diffusion coefficient, a nearly linear concentration profile is generated by means of the Soret effect~\cite{soret_1879,book_degroot_1962} in a much larger solutal diffusion time $\tau_s=L^2/D$. Since the investigated mixture has a positive separation ratio, for negative $Ra_s$ both the temperature and the concentration profile result in a stabilizing density profile~\cite{ryskin_2003} and the only variations are due to intrinsic fluctuations.

\begin{figure}[thb]
\center\includegraphics[width=8cm]{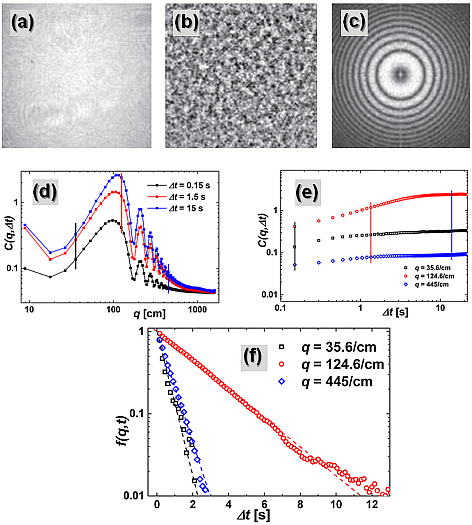}
\caption{(a) Shadowgraph image $I(\mathbf{x},t)$; (b) difference of normalized images $\Delta i_m(\mathbf{x},\Delta t)$; (c) power spectrum of (b) $\mid \Delta i_m(\mathbf{q},\Delta t)\mid ^2$; (d) structure function $C(q,\Delta t)$ for three different time delays, vertical lines stand for wave vectors used in (e); (e) structure function $C(q,\Delta t)$ for three different wave vectors, vertical lines stand for delay times used in (d); (f) ISFs for three different wave vectors $f(q,\Delta t)$: markers are for experimental data while lines depict theoretical results. All data are taken from the measurement at $Ra_s = -2\cdot10^5$.
}

\label{fig struct funct}
\end{figure}

Shadowgraphy~\cite{book_settles_2001,trainoff_2002,croccolo_2011,optical_setup} allows recording images whose intensities $I(\mathbf{x},t)$ contain a mapping of the sample refractive index fluctuations, over space and time, averaged along the direction of the gradient, as illustrated in Fig.~\ref{fig struct funct}(a). These intensity patterns are generated at the sensor plane by the heterodyne superposition of the light scattered by the sample refractive index fluctuations and the much more intense transmitted beam ('local oscillator'). These images are 2D-space-Fourier transformed \emph{in silico}, Fig.~\ref{fig struct funct}(c), to separate the contribution of light scattered at different wave vectors.
This procedure provides results similar to conventional Light Scattering, but with a shadowgraph one can access smaller wave vectors, exactly were gravity and confinement effects are expected to strongly affect the c-NEFs.

Dynamic shadowgraphy is performed by the Differential Dynamic Algorithm~\cite{croccolo_2007,croccolo_2012, croccolo_2006,cerbino_2008}, where one directly computes the so-called structure function:
\begin{align}
C(q,\Delta t)&=\langle \mid \Delta i_m(\mathbf{q},\Delta t)\mid ^2\rangle_{t,\mid  \mathbf{q}\mid=q }=\nonumber\\
&=\langle \mid i(\mathbf{q}, t)-i(\mathbf{q},t+\Delta t)\mid ^2\rangle_{t,\mid  \mathbf{q}\mid=q },
\label{eq_struct_func}
\end{align}
with $i(\mathbf{q},t)=\mathcal{F}[I(\mathbf{x},t)/\langle I(\mathbf{x},t) \rangle _{\mathbf{x}}]$ the 2D-Fourier transform of a normalized image $I(\mathbf{x},t)$ and $\Delta t$ the time delay between the pair of analyzed images, as illustrated in Fig.~\ref{fig struct funct}(b-c). $C(q,\Delta t)$ is shown in Fig.~\ref{fig struct funct}(d-e). The structure function is related to the ISF via~\cite{croccolo_2006,croccolo_2007,croccolo_2012,cerbino_2008}:
\begin{equation}
C(q,\Delta t)=2A\{T(q)S(q)\mid1-f(q,\Delta t)\mid+B(q)\},
\label{eq_SF_ISF}
\end{equation}
where $T(q)$ is the optical transfer function of the instrument (a complicated oscillating function for a shadowgraph, see~\cite{trainoff_2002,croccolo_2011}), $S(q)$ the static structure factor of c-NEFs, $A$ an intensity pre-factor, and $B(q)$ a background including all the phenomena with time-correlation functions decaying faster than the CCD frame rate, such as contributions due to shot noise and temperature fluctuations.
The ISF $f(q,{t})$ can be evaluated via Eqs.~\eqref{eq_struct_func}-~\eqref{eq_SF_ISF}. Results for three different wave vectors are shown in Fig.~\ref{fig struct funct}(f). Essentially for all the wave vectors accessible in the reported experiments the ISF can be fit by a single exponential function over the resolved part of the decay. For direct comparison with theory and simulations we extract effective decay times as the time needed to $f(q,{t})$ to decay to $1/e$.

Figure~\ref{fig times} reports experimental data for the three different $Ra_s$, not normalized in panel (a), and in dimensionless form in panel (b). For essentially all wave vectors smaller than $\tilde{q}_s^\star =\sqrt[4]{-Ra_s}$, the effective decay time departs from the theoretical description of Eq.~\eqref{eq_tau_bell} depicted as a dashed line. As the wave vector is decreased the decay time presents a minimum for a dimensionless wave vector $\tilde{q}_b\cong 5$ and for smaller wave vectors it recovers a diffusive decay $\tilde{\tau}\propto \tilde{q} ^{-2}$ (except for $Ra_s=-1\cdot10^7$, with no experimental points at low enough $\tilde{q}$).

In order to interpret these experimental findings we use a Fluctuating Hydrodynamics (FHD) model~\cite{dezarate_2006} that incorporates gravity and confinement. The dynamic structure factor $I(q,t)$ of the c-NEFs can be expressed as:
\begin{equation}
I(q,t)=S(q)f(q,t)=\sum\limits_{N=1}^{\infty}A_N(q) \exp \bigg[-\frac{t}{\tau_N(q)}\bigg],
\label{eq_dyna_struct}
\end{equation}
see~\cite{suppl} for further details. The decay times in Eq.~\eqref{eq_dyna_struct} are the inverse of the eigenvalues $\Gamma_N(q)=1/\tau_N(q)$ solving Eq.~(43) in Ref.~\cite{dezarate_2006}. The amplitudes $A_N$ are analytically related to $\Gamma_N$ and $q$. The power spectrum (static structure factor) of c-NEFs analyzed in~\cite{dezarate_2006} is then $S(q)=\sum A_N(q)$. In general, the eigenvalues can only be computed numerically, however, in the limit $q\to 0$, a full analytical investigation is possible by means of power expansions in $q$, and a clear hierarchy of well-separated $\Gamma_N$ identified~\cite{dezarate_2006}. In that limit, the first term in Eq.~\eqref{eq_dyna_struct} dominates, and $f(q\to 0,t)$ becomes single-exponential in practice, with decay time due to confinement (c):
\begin{equation}
\tilde{\tau}(\tilde{q}\to 0) |_{c} =\frac{1}{\tilde{q}^2\Big(1-\dfrac{Ra_s}{Ra_{s,c}}\Big)}=\frac{1}{\tilde{q}^2\Big(1-\dfrac{Ra_s}{720}\Big)},
\label {eq_tau_conf}
\end{equation}
where $Ra_{s,c}=720$ is the critical solutal Rayleigh number at which the convective instability first appears~\cite{ryskin_2003}. This asymptotic behaviour is shown in Fig.~\ref{fig times}(b) by dotted lines. Hence, the theory predicts a crossover from Eq.~\eqref{eq_tau_bell} (not-including confinement) at large and intermediate $q$, to the confinement behaviour of Eq.~\eqref{eq_tau_conf} at small $q$, precisely the kind of behaviour experimentally shown in Fig.~\ref{fig times}. We estimate the wave number $q_b$ corresponding to the minimum decay time by equating Eq.~\eqref{eq_tau_bell} and~\eqref{eq_tau_conf}. This gives $\tilde{q}_b=\sqrt[4]{Ra_{s,c}}=\sqrt[4]{720}\cong 5.2$ independent of $Ra_s$, in further agreement with the observations in Fig.~\ref{fig times}(b).

\begin{figure}[b]
\centering\includegraphics[width=8cm]{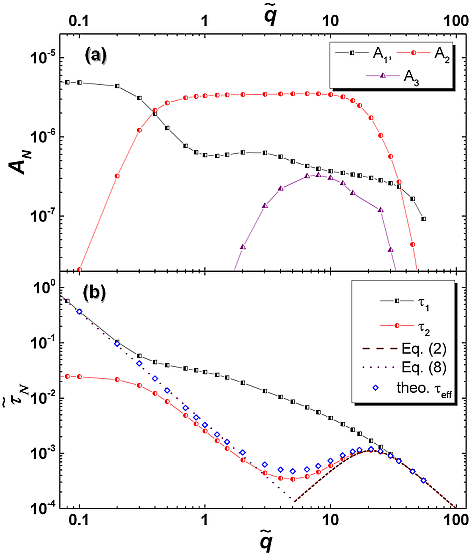}
\caption{(a) Log-log plot of the amplitude of the first three eigenmodes $A_N$, for $N=1,2,3$,  as a function of the dimensionless wave number $\tilde{q}$  for $Ra_s=-2\cdot10^5$.
(b) Log-log plot of the dimensionless decay times of the two first eigenvalues $\tilde{\tau}_N=\tau_N/\tau_s$, for $N=1,2$  as a function of the dimensionless wave number for the same $Ra_s$. Eqs.~\eqref{eq_tau_bell},~\eqref{eq_tau_conf} and the theoretical data points reported in Fig.~\ref{fig times} are also plotted for direct comparison.}
\label{fig modes}
\end{figure}

Previous work~\cite{dezarate_2006} considered only small (in magnitude) negative solutal Rayleigh numbers. Here we investigate $Ra_s$ values for realistic liquid mixtures, and find different, much richer, $\Gamma_N(q)$ and $A_N(q)$ landscapes. In Fig.~\ref{fig modes}(a) the amplitudes $A_N(\tilde{q})$ of the first three eigenmodes are shown as a function of the dimensionless wave number $\tilde{q}$, for $Ra_s=-2\cdot10^5$, while Fig.~\ref{fig modes}(b) shows the corresponding dimensionless decay times $\tilde{\tau}_N(\tilde{q})$ for the first two modes. Clearly in different wave number ranges different modes dominate. For very large ($\tilde{q} \gtrsim 50$) wave numbers, all decay times collapse and the ISF is approximately a single exponential dominated in amplitude by the first mode. For very small wave numbers ($\tilde{q}\lesssim0.3$) the first mode dominates in amplitude and a single-exponential decay is again recovered. The second mode leads in amplitude in the central range $0.6 \lesssim \tilde{q}\lesssim30$, but having a smaller decay time means that both modes play a significant role and the ISF should show signatures of a double exponential decay. Indeed, data from theory and simulations show such signatures in the predicted wave vector range, however such signatures are not detectable in the experimental data given the limited wave vector range and the smaller signal to noise ratio. In Fig.~\ref{fig struct funct}(e) we report three examples of experimental ISFs for different wave vectors.

Also for the theory and regardless of the multiple exponential character, a single effective decay time $\tau_{\text{eff}}(q)$ is defined by $f(q,\tau_{\text{eff}})=1/e$. In Figs.~\ref{fig times} and~\ref{fig modes}(b) we show results for $\tau_{\text{eff}}(q)$, computed via Eq.~\eqref{eq_dyna_struct} from the amplitudes and decay rates obtained theoretically. All the features seen in the experimental data points are well-reproduced by the theory. Noticeably the slowing-down observed for small wave numbers is clearly related to confinement, since this is the only ingredient added to the theory that gives Eq.~\eqref{eq_tau_bell}.

The theory~\cite{dezarate_2006} assumes that viscous dissipation dominates, and neglects the effect of fluid inertia; this is justified by the fact that in all liquids momentum diffusion is much faster than mass diffusion, i.e., the Schmidt number $\text{Sc}=\nu/D$ is very large. While neglecting inertial effects is a good approximation at most wavenumbers of interest, it is known that, depending on $Ra_s$, it fails at sufficiently small wavenumbers due to the appearance of inertial \textit{propagative modes}~\cite{takacs_2008} (closely related to gravity waves) driven by buoyancy.
In order to confirm that the observed slowing down is due to confinement and not to inertia we have performed numerical simulations that account for inertial effects and confinement~\cite{usabiaga_2012,delong_2014}, see~\cite{suppl} for details. Data points from a numerical simulation with fluid parameters matching the experimental ones are also shown in Fig.~\ref{fig times}. An excellent agreement is visible for this dataset among experimental, theoretical and simulation results, confirming that inertia effects are not relevant in our experiments. We note, however, that for thicknesses $L \gtrsim 5$ mm the simulations do show oscillatory time correlation functions (propagative modes) at the smallest wavenumbers~\cite{delong_2014}; this range is not accessible in the experiments reported here.


We conclude that confinement has a moderate damping impact on the intensities of large-scale non-equilibrium concentration fluctuations, but, in the presence of gravity, it strongly affects their dynamics.  Experiments, as well as theory and simulations, show that the slowing down is determined by the solutal Rayleigh number that, in this study, is controlled solely by the confinement distance $L$. Large-scale fluctuations are confined to evolve in an essentially quasi-two-dimensional manner by the boundaries, and we found them to behave diffusively but with a greatly enhanced  diffusion coefficient.

This is in contrast to the case of diffusion in a microgravity environment where the coupling to velocity fluctuations greatly enhances the intensity of the c-NEFs but does not alter their Fickian diffusive dynamics~\cite{vailati_2011}. In strongly confined systems, such as porous media, the buoyancy driven acceleration of the fluctuations is eliminated by confinement already at mesoscopic scales. In the absence of confinement, however, the gravitational acceleration is eventually suppressed by inertial effects leading to propagative modes; confinement is expected to also strongly affect the dynamics of these gravity waves, as we will explore in future work.

Although the main focus of this letter is on the dynamics and we leave for future publications a full discussion of the statics, we note that the minimum $\tilde{q}_b$ in $\tau_\text{eff}$ corresponds to a minimum in the intensity of fluctuations $S(q)$. This indicates that the results presented here may be thought of as a kind of de Gennes narrowing~\cite{degennes_1959}. In analogy to diffusion in colloidal suspensions where a competition between interparticle interactions and hydrodynamic effects is present, here we have a competition between gravity and confinement.

Interestingly, we find that the dimensionless wave number identifying where confinement coexists with gravity is related to the critical solutal Rayleigh number $Ra_{s,c}=720$ where the convective instability first appears~\cite{ryskin_2003}. This is a signature of the Onsager regression hypothesis stating that the dynamics of the fluctuations contains all of the signatures seen in the deterministic dynamics, which is known to be controlled by the Rayleigh number.  Our work indicates that the study of the dynamics, rather than the intensity of non-equilibrium fluctuations, gives deep insights into the competition of physical processes such as diffusion, buoyancy, and confinement.

\begin{acknowledgments}
J.M.O.d.Z. acknowledges support from the UCM/Santander Research Grant PR6/13-18867 during a sabbatical leave at Anglet, when part of this work was developed. F.C. acknowledges fruitful discussions with Alberto Vailati, Doriano Brogioli and Roberto Cerbino. A.D. was supported in part by the U.S. National Science Foundation under grant DMS-1115341 and the Office of Science of the U.S. Department of Energy through Early Career award number DE-SC0008271.\newline
\end{acknowledgments}

\indent Correspondence and requests for materials should be addressed to F.C. (fabrizio.croccolo@univ-pau.fr)

\end{document}


\title{Supplemental material for:\\ Slowing-down of non-equilibrium concentration 
fluctuations in confinement.}
\author{C\'edric Giraudet$^{1}$, Henri Bataller$^{1}$, Yifei Sun$^{2}$, Aleksandar Donev$^{2}$, Jos\'e Maria Ortiz de Z\'arate$^{3}$ and Fabrizio Croccolo$^{1}$}

\address{$^{1}$Laboratoire des Fluides Complexes et leurs R\'eservoirs, Universit\'e de Pau et des Pays de l'Adour, 64600 Anglet, France}
\address{$^{2}$Courant Institute of Mathematical Sciences, New York University, New York, NY 10012, USA}
\address{$^{3}$Departamento de F\'{\i}sica Aplicada I, Universidad Complutense, 28040 Madrid, Spain}

\date{\today}

\maketitle

\renewcommand{\theequation}{S\arabic{equation}}
\renewcommand{\thefigure}{S\arabic{figure}}

\section*{Fluctuating Hydrodynamics}

Thermal fluctuations in non-equilibrium systems can be described with the method of fluctuating hydrodynamics (FHD) \cite{book_dezarate_2006}. For the purpose of this letter, we consider the FHD equations of a binary mixture in the presence of a stationary concentration gradient and in the limit of large Lewis number~\cite{dezarate_2006,velarde_1972} and large Schmidt number~\cite{donev_2014}. This approximation is adequate for liquid mixtures (but not for gases) of positive separation ratio \cite{negative_sp} and, in the presence of gravity, for sufficiently large wave vectors. By linearizing the complete hydrodynamics equations to leading order in the fluctuations one obtains for the small fluctuations around the steady state~\cite{dezarate_2006}:
\begin{equation}
\begin{split}
0&=\mathbf{\nabla} ^4 \delta v_z -\beta_s g (\partial^2_x+\partial^2_y)\delta c+\frac{1}{\rho} \{ \nabla\!\times\!\nabla\!\times\!(\nabla\!\cdot\!\Pi) \}_z
\\
\partial_t(\delta t)&=D \nabla^2(\delta c)-(\nabla c)\delta v_z,
\raisetag{12pt}\label{eq_system_v_c}
\end{split}
\end{equation}
where $\delta v_z$ are the fluctuations in the fluid velocity component along the direction of the gradient, $\nabla c=\boldsymbol{\nabla}c\cdot\hat{z}$, and $\delta c$  is the mass concentration fluctuations. In Eqs.~\eqref{eq_system_v_c}, in accordance to the general guidelines of FHD, we added a white-noise stochastic momentum flux, $\Pi(\mathbf{r},t)$, whose statistical properties are given by the fluctuation-dissipation theorem~\cite{book_dezarate_2006,dezarate_2006}. All other symbols are defined after Eq.~(1) of the main text.

The problem of FHD is to solve Eqs.~\eqref{eq_system_v_c} so as to obtain the correlation function of concentration fluctuations (proportional to the measured ISF \cite{book_dezarate_2006,croccolo_2012}) from the correlation function of the stochastic noise $\Pi(\mathbf{r},t)$. If one does not consider boundary conditions for the fluctuating fields, Eqs.~\eqref{eq_system_v_c} are readily solved in the Fourier domain \cite{segre_1993a,segre_1993b}, leading to a single exponential decay of the ISF with dimensionless decay time given by Eq.~(2) of the main text. This solution without boundary conditions is only meaningful for negative $Ra_s$. To account for confinement effects, one has to implement realistic boundary conditions:
\begin{equation}
0=\delta v_z=\partial_z \delta v_z=\partial_z \delta c \quad \text{at} \quad z=0 \text{ and } z=L.
\label{eq_bound_cond}
\end{equation}
The linear stability of the  problem given by Eqs.~\eqref{eq_system_v_c}-\eqref{eq_bound_cond} was studied by Ryskin \textit{et al.} \cite{ryskin_2003} who showed that the quiescent state is stable below a convection threshold: $Ra_s<Ra_{s,c}=+720$.

In a previous publication of theoretical nature \cite{dezarate_2006}, it was shown how conditions~\eqref{eq_bound_cond} can indeed be implemented in FHD by expanding the solution of Eqs.~\eqref{eq_system_v_c} in a series of hydrodynamic modes that solve an associated eigenvalue problem. The hydrodynamic modes (eigenfunctions) have single decay times $\tau_N(q)$ that are obtained by numerically solving an algebraic equation (Eq.(43) in Ref.\cite{dezarate_2006}). Although all the data reported in this letter are for negative $Ra_s$, we note that the solution to Eqs.~\eqref{eq_system_v_c} incorporating the boundary conditions~\eqref{eq_bound_cond} is meaningful for any $Ra_s<+720$, including the whole range of positive solutal Rayleigh numbers below the convection threshold \cite{dezarate_2006,ryskin_2003}.

The focus of \cite{dezarate_2006} was on the static structure factor $S(q)$ of c-NEFs and the dynamics, although implicitly included, was not investigated. In this work we focus on the dynamics, and obtain Eqs.~(5) and~(6) of the main text. In addition, we numerically evaluate the decay times $\tau_N(q)$ and corresponding amplitudes $A_N(q)$ for a range of previously unexplored large negative values of the solutal Rayleigh number relevant to the experimental conditions. The results of these calculations for $Ra_s = -2\cdot 10^5$ are summarized in Fig.~4 of the main text. From the data contained in this figure, using Eq.~(5), we computed theoretical ISFs that, depending on the wave number $q$, exhibit a clear multi-exponential behaviour that will be discussed more in detail in future publications. A well-defined effective decay time $\tau_{\text{eff}}$ can be extracted by evaluating the time the ISF takes to decay to $1/e$. Those theoretical decay times are displayed in Figs.~3(b) and~4 as open blue symbols, and compared with experiments and simulations.

\section*{Numerical simulations}

We perform computer simulations of the experimental setup using finite-volume methods for fluctuating hydrodynamics described in more detail elsewhere \cite{usabiaga_2012,delong_2014,donev_2014}; here we summarize some key points. The numerical methods have been implemented in the IBAMR software framework \cite{griffith_2007}. The numerical codes solve the following stochastic partial differential equations for the fluctuating velocity field $\mathbf{v}(\mathbf{r},t)$ and mass concentration $c(\mathbf{r},t)$ in a binary liquid mixture \cite{book_dezarate_2006},
\begin{align}\label{eq_system_simul1}
&\rho\partial_t\mathbf{v}+\nabla p=\eta \nabla^2 \mathbf{v}+\nabla\!\cdot\!\Pi - \rho \beta_s c \mathbf{g}\\
&\nabla\!\cdot\!\mathbf{v}=0\nonumber\\
&\partial_t(c)+\mathbf{v}\!\cdot\!\nabla c=D \nabla^2c,
\label{eq_system_simul2}
\end{align}
where $\eta=\rho\nu$  is the shear viscosity, $p(\mathbf{r},t)$ the pressure and, as in Eqs.~\eqref{eq_system_v_c}, $\Pi(\mathbf{r},t)$ denotes a white-noise stochastic momentum flux due to thermal fluctuations.
Nonlinear advective terms in the velocity equation are neglected in a small Reynolds number approximation. Temperature fluctuations are not considered in a large Lewis number (very fast temperature dynamics) approximation \cite{dezarate_2006}. We assume that the applied temperature gradient $\nabla{T}$ is weak and approximate  $T\approx T_o=298$ K. Note that in~\eqref{eq_system_simul2}, as well as in~\eqref{eq_system_v_c}, we have ignored thermal fluctuations in the mass flux, which are responsible for equilibrium fluctuations in the concentration. This is because our focus is on the much larger non-equilibrium fluctuations induced by the coupling to the velocity equation via the advective term $\mathbf{v}\!\cdot\!\nabla c$.

In linearized FHD the equations~\eqref{eq_system_simul1}-\eqref{eq_system_simul2} are expanded to leading order in the magnitude of the fluctuations $\delta c=c-\langle c \rangle$ and $\delta \mathbf{v}=\mathbf{v}-\langle \mathbf{v} \rangle$ around the steady state solution of the deterministic equations \cite{dezarate_2006}. Typical liquid mixtures have a large Schmidt number, $Sc=\nu/D\gg 1$, and under certain conditions one can take a limit of equations~\eqref{eq_system_simul1}-\eqref{eq_system_simul2} as $Sc\to \infty$; in the linearized setting this over-damped limit amounts to deleting the inertial term $\rho \partial_t \mathbf{v}$ in the velocity equation~\eqref{eq_system_simul1},~\cite{dezarate_2006,delong_2014,ryskin_2003,velarde_1972}. With these simplifications only the component of the velocity parallel to the macroscopic concentration gradient couples to the concentration equations, and after taking a double curl of the velocity equation one obtains equations~\eqref{eq_system_v_c},~\cite{dezarate_2006}.
Our numerical method, however, solves the complete hydrodynamic equations~\eqref{eq_system_simul1}-\eqref{eq_system_simul2}  in two dimensions with the concentration gradient along the $y$ axis; for this problem there is no difference between two and three-dimensional simulations due to the symmetries of the problem.

With a simple modification of the time-integration algorithm used in the numerical method we can perform simulations with or without the $\rho \partial_t \mathbf{v}$ term in the velocity equation~\eqref{eq_system_simul1}, allowing us to study the importance of fluid inertia \cite{delong_2014}. In the inertia-less limit we have confirmed that numerical simulations reproduce the results of the theoretical calculations (see previous section on FHD) based on solving~\eqref{eq_system_v_c}-\eqref{eq_bound_cond} analytically. In the simulations with inertia we have confirmed that, for the range of $Ra_s$ probed experimentally, propagative modes appear only for the largest $Ra_s\!=\!10^7$ but at wavenumbers $q\lesssim10$ cm$^{-1}$ not resolved in the experiments. For the wavenumbers and Rayleigh numbers experimentally studied, simulations show negligible effects of inertia on the correlation functions, and the same slow decay at long times is observed for confined fluctuations with or without fluid inertia. Hence, the observed slowing-down is due to confinement only.

In our simulations, the domain is periodic along the $x$ direction. At the top and bottom boundaries, $y=0,L$,  a no-flux boundary condition is imposed for the concentration, $\nabla c=-c_o(1-c_o)S_T\nabla T$, where $c_o=0.5$  is the average concentration, and a no-slip boundary condition is imposed for velocity. For comparison, we have also performed simulations employing free-slip boundary conditions for the velocity; these show a qualitatively similar behaviour to the results reported here but differ quantitatively indicating the importance of the boundary conditions (confinement).

The experimentally observed light intensity, once corrected for the optical transfer function of the equipment, is proportional to the intensity of the fluctuations in the concentration averaged along the gradient \cite{book_dezarate_2006,dezarate_2006,trainoff_2002},
\begin{equation}
\delta c_{\perp}(x;t)=\frac{1}{L}\int_0^L \!\delta c (x,y;t) dy.
\label{eq_integral}
\end{equation}

The main quantity of interest in our simulations is the Fourier transform $\delta \hat{c}_{\perp}(q;t)$ of the vertically averaged concentration, the time-correlation of which gives the dynamic structure factor appearing in Eq.~(5) of the main text:
\begin{equation}
I(q,t)=\langle \delta \hat{c}_{\perp}(q,t) \delta \hat{c}_{\perp}^\star (q,0)\rangle.
\label{eq_spectrum}
\end{equation}
After suitable normalization and background subtraction $I(q,t)$ is directly related to the experimental ISF, see Eq.~(4) of the main text. We obtain the relaxation time $\tau$ from the simulation results by fitting $I(q,t)$ to a sum of two exponentials and solving $I(q,\tau)=I(q,0)/e$.

The physical parameters used in the simulations are the same as reported in Ref. [23] of the main text and the temperature difference across the sample is  $\Delta T=20$~K, heating from the top boundary. The length of the simulation box in the periodic direction (perpendicular to the gradient) is 6.13 mm. The time step size is sufficiently small to resolve the fast viscous dynamics, $\Delta t\!=\!5 \times 10^{-3}$~s. We skip the initial 1250 seconds (in physical time) of the run to allow the steady state to develop, and then collect data for another 6250 seconds. Different sizes of uniform grids were used for the different sample thicknesses, as summarized in the following table:

\[
\begin{tabular}{|c|c|c|}
\hline
Thickness, $L$ & ${Ra}_{s}$ & grid resolution\tabularnewline
\hline
\hline
0.7~mm & $-4\cdot10^{4}$ & $280\times32$\tabularnewline
\hline
1.3~mm & $-2\cdot10^{5}$ & $300\times64$\tabularnewline
\hline
5~mm & $-1\cdot10^{7}$ & $156\times128$\tabularnewline
\hline
\end{tabular}
\]

\section*{Phenomenological equation for the decay times including: diffusion, gravity and confinement}

We now focus on the shape of the dimensionless decay time $\tilde{\tau}$ as a function of $\tilde{q}$, with the goal of obtaining an empirical relation to replace Eq.~(2) of the main text and incorporate the confinement as described by Eq.~(6)  for small wave numbers. Specifically, we propose a phenomenological equation that crosses over analytically from Eq.~(6) of the main text at extremely small $\tilde{q}$ to Eq.~(2) at intermediate $\tilde{q}$, and is therefore useful for a rapid analysis of experimental results.
We first notice that the only difference between the two aforementioned equations is that the $\tilde{q}^4$ term of Eq.~(2) becomes 720 in Eq.~(6), so we substitute these terms by $720(1+\tilde{q}^4/720)$, to obtain the empirical equation:
\begin{align}
\frac{\tau(\tilde{q})}{\tau_s} \bigg|_{d+g+c} &=\frac{1}{\tilde{q}^2\Bigg[1-\dfrac{Ra_s}{720(1+\frac{\tilde{q}^4}{720})}\Bigg]}=\nonumber\\
&=\frac{720+\tilde{q}^4}{\tilde{q}^2\big(-Ra_s+720+\tilde{q}^4\big)}.
\label{eq_tau_all}
\end{align}
\begin{figure}[t]
\center\includegraphics[width=8cm]{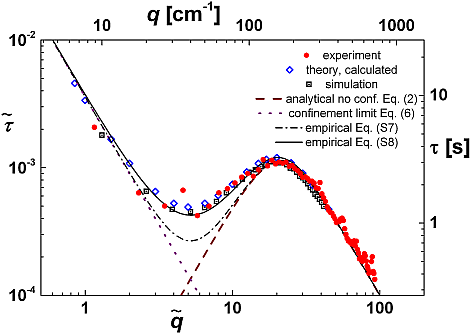}
\caption{Log-log plot of the experimental decay times as a function of wave vector for $Ra_s=-2\cdot10^5$.
Filled red circles are experimental data, open blue diamonds for calculations based on the FHD model, and open-dotted black squares from numerical simulations. The brown dashed line depicts the analytical solution provided by Eq.~(2), taking into account gravity and diffusion only; the purple dotted line stands for the confinement limit of Eq.~(6). The dashed-dotted black line represents the empirical formula expressed in Eq.~\eqref{eq_tau_all}, while the continuous black line is for Eq.~\eqref{eq_tau_alpha} with $\alpha = 32$. }
\label{fig empirical}
\end{figure}
We show in Fig.~\ref{fig empirical} the crossover curve, Eq.~\eqref{eq_tau_all}, together with the experimental, theoretical and simulation data points for $Ra_s=-2\cdot10^5$. Clearly a discrepancy is present between Eq.~\eqref{eq_tau_all} and the data points for wave vectors around $q_b\!=\!5.2$. The gap can be filled-in by slightly modifying Eq.~\eqref{eq_tau_all}, adding a term of the second order in $\tilde{q}$ in the numerator:
\begin{equation}
\frac{\tau(\tilde{q})}{\tau_s} \bigg|_{d+g+c} =\frac{720+\alpha \tilde{q}^2+\tilde{q}^4}{\tilde{q}^2\big(-Ra_s+720+\tilde{q}^4\big)},
\label{eq_tau_alpha}
\end{equation}
which does not modify the asymptotic behaviours at large and small $\tilde{q}$. We find that Eq.~\eqref{eq_tau_alpha} can be used in practice to fit the experimental data points with the \textit{ad hoc} constant $\alpha$ as a free parameter.